\documentclass[fleqn,10pt]{wlscirep}
\usepackage[utf8]{inputenc}
\pdfoutput=1

\usepackage[T1]{fontenc}
\usepackage{algorithm}
\usepackage{algorithmic}
\usepackage{dirtytalk}

\usepackage{doi}
\usepackage{graphicx}
\usepackage{multicol}
\usepackage{multirow} 

\def\BibTeX{Bib\TeX}
\def\BibTeX{{\rm B\kern-.05em{\sc i\kern-.025em b}\kern-.08em
    T\kern-.1667em\lower.7ex\hbox{E}\kern-.125emX}}

\usepackage{xcolor}

\usepackage{cite}
\usepackage{amsmath,amssymb,amsfonts,color,colortbl,textgreek}

\usepackage{lettrine,paralist,fancyhdr}
\usepackage{nomencl,makeidx,pdfpages}
\usepackage{setspace}

\usepackage{longtable}
\usepackage{booktabs} 

\usepackage{bigdelim} 
\usepackage{bigstrut} 
\usepackage{tabularx}
\usepackage{ragged2e}
 \usepackage{float}

\usepackage{textcase}
\usepackage{listings}

\usepackage{pgfplots}
\usepackage{pgfplotstable}
\usepackage{tikz}
\pgfplotsset{compat=newest}
\pgfplotsset{compat=1.18}
\usepackage{eqparbox}

\usepackage{siunitx}

\usepackage[labelfont=bf]{caption}
\usepackage{subcaption}
\usepackage{varioref}

\usepackage[numbers]{natbib} 
\setlist[itemize]{itemsep=-5pt, topsep=0pt} 

\usepackage{hyperref} 
\usepackage{orcidlink} 

\title{Scientific Reports Title to see here}

\title{Design Multiband Monopole and Microstrip Patch Antennas using High Frequency Structure Simulator}
\author[1]{Georgios Giannakopoulos\orcidlink{0000-0002-3707-3276}}
\author[2]{Khushbu Mehboob Shaikh\orcidlink{0009-0000-8681-5830}}

\affil[1]{Independent Researcher, The Hague, The Netherlands} 
\affil[2]{Technical Lead, Staff Technical Account Manager, Twilio Inc., Irving, Texas, United States} 


\begin{abstract}
This paper describes the design, construction and testing of a dual-band monopole microstrip patch antenna to operate in the GSM and DCS bands. For compactness the monopole antenna is mounted on top of a FR4 substrate and is designed as a planar structure which can be wrapped into a box like structure. The performance characteristics of the wrapped dual band monopole antenna are simulated on Ansoft HFSS before fabrication and testing on a Network Analyzer. The simulated HFSS return loss of the antenna shows broad agreement with experiment over the frequency range from \(500 ~\mathrm{MHz} \text{ to } 4~\mathrm{GHz}\). Radiation patterns are generated in each band of operation of the antenna. The effect of feedline impedance matching on the performance of the monopole antenna is discussed.
\end{abstract}
\begin{document}

\flushbottom
\maketitle
%
%
\thispagestyle{empty}


\section{Introduction}
\label{section1}
This paper is concerned with the design, construction and testing of a dual-band monopole antenna which is required to operate in the Global System for Mobile communications (GSM \(890 - 960 \mathrm{MHz}\)) and Digital Communication System (DCS \(1710 - 1880 \mathrm{MHz}\)) bands. Monopole antennas have similar characteristics to a dipole antenna in which the monopole structure can be represented by one half of the dipole and the ground plane of the monopole can be represented by the remaining half of the dipole. The fundamental resonant frequency of a monopole is therefore related to a quarter wavelength of a patch length in the monopole structure. The principle involved in achieving a multiband antenna is for the structure of the monopole to be made up of multiple path lengths with each length being associated with a quarter wavelength of each resonant frequency. In practice, for compact monopole structures, the resonant frequencies do not occur exactly at a quarter wavelength of each path since coupling effects occur between the close multiple paths of the microstrip patches. For compactness and efficiency, mobile phone dual-band antennas are constructed from conductive patches being placed on a microstrip antenna. \\

\noindent
Microstrip antennas have received much attention over recent years due to their low cost, robustness, ease of construction and compact size. These antennas can be mounted on aircraft, spacecraft, satellites, cars and mobile telephones. Microstrip antennas received considerable attention in the 1970’s but was first patented in 1955 \cite{Gutton1955}. These antennas basically consist of a conductive strip (patch) and a conductive ground plane which is separated by a thin substrate of dielectric material. Popular feeding methods to the antenna are the microstrip line, coaxial probe, aperture coupling and proximity coupling. The patch of the antenna can take on a variety of different shapes. However the theory involved is complex and approximate theories have only been developed for simple shapes such as rectangular or circular patches. \\

\noindent
In this paper, use of the Ansoft High Frequency Structure Simulator (HFSS) \cite{Ansoft2008} package has been used to determine the performance characteristics of a GSM/DCS dual-band monopole antenna. The return loss results are compared with experimental values obtained from a fabricated antenna connected to a network analyzer.

\section{Dual Band Monopole Design}
\label{section2}
A monopole antenna is an antenna which is similar to a dipole antenna but with one half of the dipole being replaced by a ground plane at right angles to the other half of the dipole. When the ground plane is large, a monopole antenna behaves like a dipole antenna. A vehicle radio aerial for example is a monopole antenna in which one half of the dipole antenna is the mast and the other half of the dipole antenna is the body of the vehicle.\\
\noindent
Mobile phones also use monopole antennas. In the past, mobile phones had a protruding aerial acting as one half of the dipole. The disadvantage of a protruding monopole was that the size of the phone was increased and also they were susceptible to breaking off. Modern mobile phones have a monopole antenna in which is enclosed within the body of the phone in the form of a compact wrapped structure.
Mobile phones also are required to produce low radiation towards a user’s head. Such phones are said to produce a low specific absorption rate (SAR). Low profile monopole antennas which incorporate folded elements as a wrapped structure does block some of the radiation towards a user’s head.\\
\noindent
In recent years, mobile phones are required to receive multi-band signals is which the antenna must resonate at predetermined frequencies. For example a dual-band phone which receives GSM/DCS is required to have two separate resonant bandwidths in operating around \(900 \mathrm{MHz}\) and \(1800 \mathrm{MHz}\) frequencies. A triple band phone which receives GSM/DCS/PCS is required to have three separate resonant bandwidths operating at frequencies of \(900 \mathrm{MHz}\), \(1800 \mathrm{MHz}\) and \(1900 \mathrm{MHz}\) \cite{Teng2002}. In order for mobile phones to be compact in size, planar monopoles operating at these frequencies need be bent or folded into certain shape. For example a long quarter wave path can be compacted from a \(2 \mathrm{D}\) planar structure into a \(3 \mathrm{D}\) structure by spiralling the patch inwards or outwards. Antennas can also be wrapped into \(3 \mathrm{D}\) structures by bending the quarter wave paths into suitable shapes \cite{Chang2002}, \cite{Lee2002b}, \cite{Lee2002c}.\\

\begin{figure}[ht]
    \centering
    \begin{subfigure}[b]{0.68\textwidth}
        \centering
        \includegraphics[width=\textwidth]{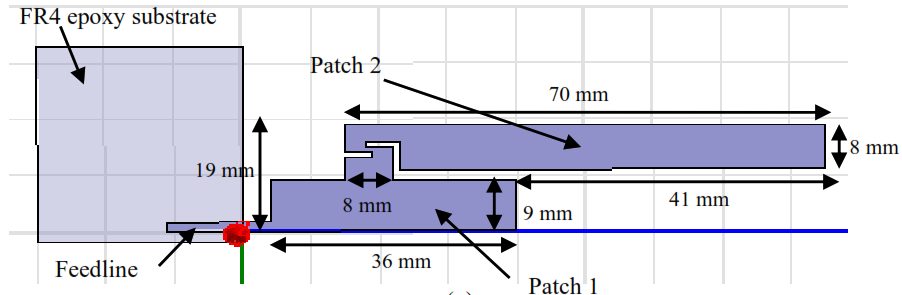}  
        \caption*{(a)}
        \label{fig:(a)}
    \end{subfigure}
    \hfill
    \begin{subfigure}[b]{0.68\textwidth}
        \centering
        \includegraphics[width=\textwidth]{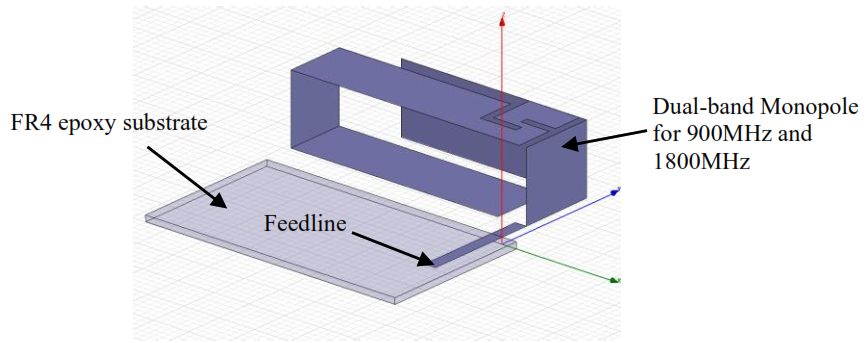}   
        \caption*{(b)}
        \label{fig:(b)}
    \end{subfigure}
   \caption{ Monopole antennas layout (a) unwrapped and (b) wrapped.}
    \label{fig:1}
\end{figure}

\noindent
In this paper Ansoft HFSS is used to design a dual-band monopole antenna for GSM \((890 - 960 \mathrm{MHz})\) and DCS \((1710 - 1880 \mathrm{MHz})\) operation.
The dual-band antenna discussed in this paper is shown in 
\hyperref[fig:1]{Figure 1 (a)} and \hyperref[fig:1]{Figure 1 (b)}. 

\noindent
 \hyperref[fig:1]{Figure 1 (b)} shows the wrapped version of the two dimensional form of the antenna shown in \hyperref[fig:1]{Figure 1 (a)}. The wrapped structure in \hyperref[fig:1]{Figure 1 (b)} is a compact structure which has overall dimensions of   and is therefore a suitable design for mobile phone use.\\
\noindent
The monopole of the antenna is surrounded by air and is attached to a feedline which passes over an FR4-epoxy substrate which has a thickness of \(1.6 mm\) and a relative permittivity value of \(4.7\). The feedline is extended by a distance of \(4 mm\) away from the end of the substrate so that coupling effects between the monopole and the ground plane below the substrate are reduced.

\section{HFSS results}
\label{section3}
\begin{figure}[h]
\begin{center}
{
\includegraphics[width=0.70\textwidth]{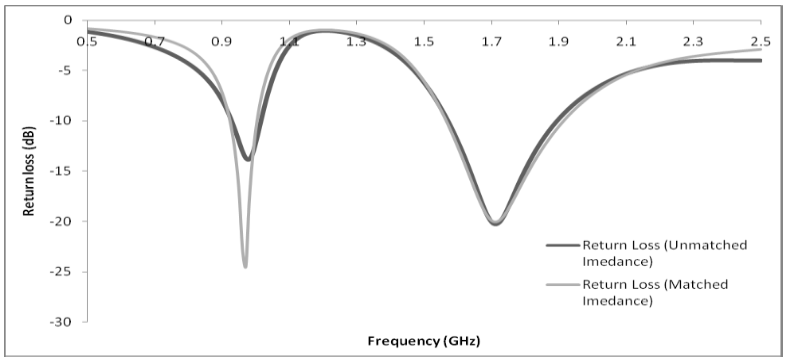}
}  
         \caption{ Microstrip Antenna under test on a Network Analyzer.}
         \label{fig:2}
          \end{center}   
\end{figure}

The monopole structure in \hyperref[fig:1]{Figure 1 (b)} has been simulated in Ansoft HFSS over a frequency range from  \(500 ~\mathrm{MHz} \text{ to } 4~\mathrm{GHz}\). The feedline is designed to have an impedance of \(50 \Omega\) which requires the thickness of the feedline strip to be \(2.91 mm\). The length of the feedline used in the simulation was chosen to be \(100 mm\) in order to avoid any wave port radiation contaminating the results. The return loss for the wrapped structure shown in \hyperref[fig:1]{Figure 1 (b)} is presented in \hyperref[fig:2]{Figure 2}, using both a standard feedline (unmatched impedance) and a feedline attached to a quarter wave transformer (matched impedance at a frequency of \(980 \mathrm{MHz}\)).

\begin{figure}[h]
\begin{center}
{
\includegraphics[width=0.70\textwidth]{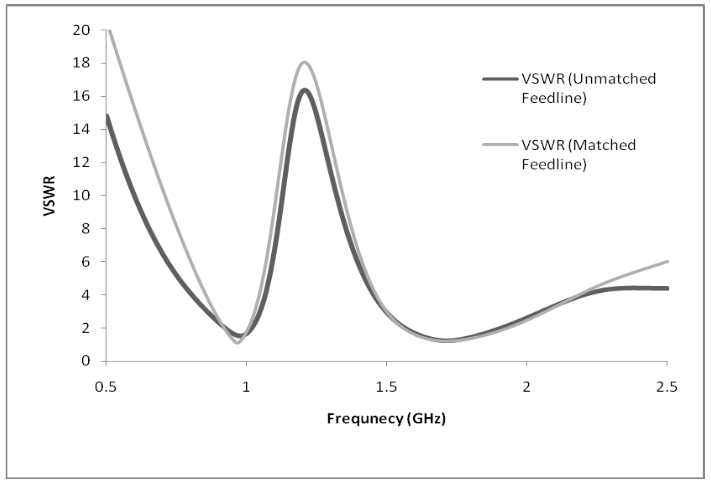}
}  
         \caption{Return Loss (dB) plotted against frequency (GHz) 
 (a) Unmatched feedline impedance (b) Matched feedline impedance}
         \label{fig:3}
          \end{center}   
\end{figure}

\noindent
\hyperref[fig:2]{Figure 2} shows two resonant frequency bands which operate in the GSM band and the DCS band. The unmatched feedline return data show a return loss of \(13.9 dB\) at a resonant frequency of \(980 \mathrm{MHz}\) and a return loss of \(20.3 dB\) at a resonant frequency of \(1.71 \mathrm{GHz}\). There is a considerable improvement in the gain of the antenna at a resonant frequency of \(980 \mathrm{MHz}\) when a matched impedance feedline is used and no noticeable gain difference in the return loss at the resonant frequency of \(1.71 \mathrm{GHz}\).\\
\noindent
The VSWR values corresponding to the return loss data in \hyperref[fig:2]{Figure 2} is presented in \hyperref[fig:3]{Figure 3}.

\noindent
An efficient antenna should have a VSWR value of less than \(1.5\) at any particular resonant frequency of interest. A VSWR value of \(2\) is obtained with an unmatched impedance feedline and a VSWR value of \(1.5\) is achieved for a matched impedance feedline. A VSWR value of \(2\) is obtained at \(1.71 \mathrm{GHz}\) for both matched and unmatched feed lines. It should be noted that a lower VSWR value could have been achieved at the resonant frequency\(1.71 \mathrm{GHz}\) by using a different feedline transformer which is matched at the frequency of \(1.71 \mathrm{GHz}\) instead of the frequency \(980 \mathrm{MHz}\).\\

\begin{figure}[ht]
    \centering
    \begin{subfigure}[b]{0.40\textwidth}
        \centering
        \includegraphics[width=\textwidth]{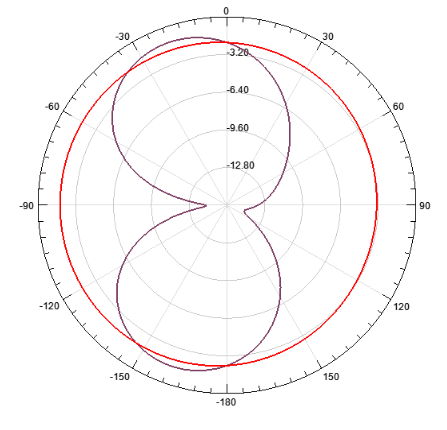}  
        \caption*{(a)}
        \label{fig:(a)}
    \end{subfigure}
    \hfill
    \begin{subfigure}[b]{0.40\textwidth}
        \centering
        \includegraphics[width=\textwidth]{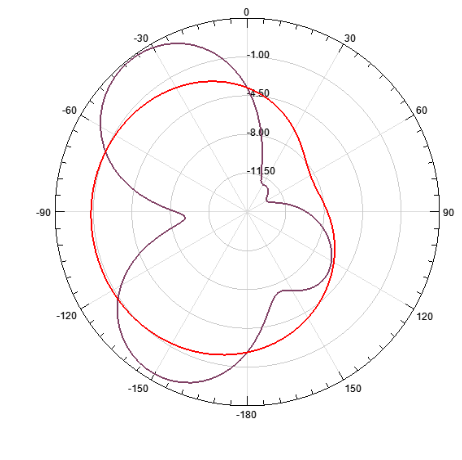}   
        \caption*{(b)}
        \label{fig:(b)}
    \end{subfigure}
   \caption{Radiation pattern for wrapped layout at (a) \(890\mathrm{MHz}\) and (b) \(1710 \mathrm{MHz}\).}
    \label{fig:4}
\end{figure}

\noindent
The radiation patterns on the \(xz-\)plane (corresponding to the plane \(\phi=0^\circ\)) and \(yz-\)plane (corresponding to the plane \(\phi=90^\circ\)) are shown in \hyperref[fig:4]{Figure 4}. The radiation pattern at \(980 \mathrm{MHz}\), shown in \hyperref[fig:4]{Figure 4 (a)}, is a typical dipole radiation pattern at its fundamental frequency mode of operation. There is a constant radiation pattern in the \(xz-\)plane and a radiation pattern containing circular lobes in the \(yz-\)plane. The radiation pattern in the DCS band shown in \hyperref[fig:4]{Figure 4 (b)} is more distorted than the GSM band leading to larger variations in the gain in different directions.\\

\begin{figure}[!h]
\begin{center}
{
\includegraphics[width=0.60\textwidth]{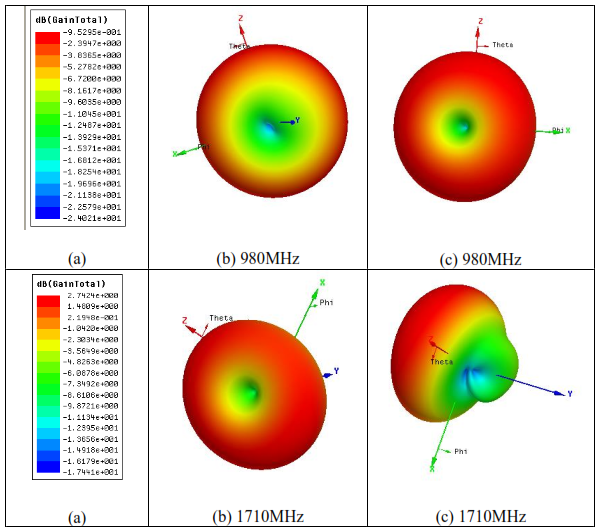}
}  
         \caption{Three dimensional radiation pattern at \(980 \mathrm{MHz}\) and \(1710 \mathrm{MHz}\) resonant frequencies (a) dB Gain Colour scales, (b) radiation pattern from front side (c) radiation pattern from back side. }
         \label{fig:5}
          \end{center}   
\end{figure}

\noindent
\hyperref[fig:5]{Figure 5} shows the corresponding three dimensional form of the far field radiation pattern at \(980 \mathrm{MHz}\) and \(1710 \mathrm{MHz}\) frequencies. Rotated views of the patterns in \hyperref[fig:5]{Figure 5 (b)} are shown in \hyperref[fig:5]{Figure 5 (c)}.

\newpage 
\section{Fabrication and Experimental Results}
\label{section4}
\begin{figure}[ht]
    \centering
    \begin{subfigure}[b]{0.48\textwidth}
        \centering
        \includegraphics[width=\textwidth]{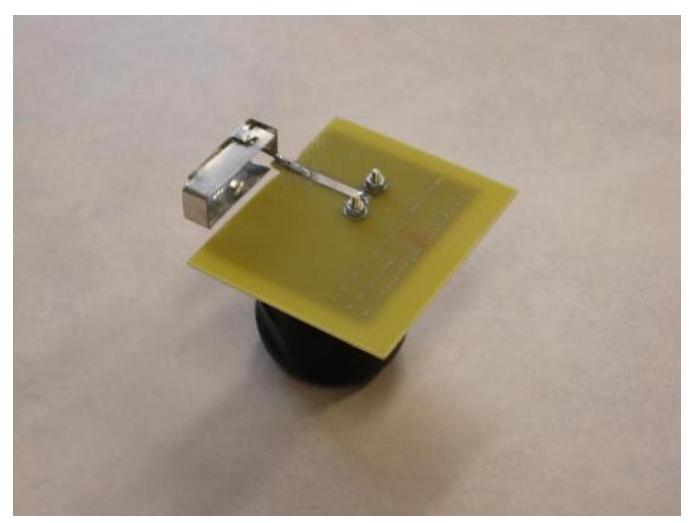}  
        \caption*{(a)}
        \label{fig:(a)}
    \end{subfigure}
    \hfill
    \begin{subfigure}[b]{0.48\textwidth}
        \centering
        \includegraphics[width=\textwidth]{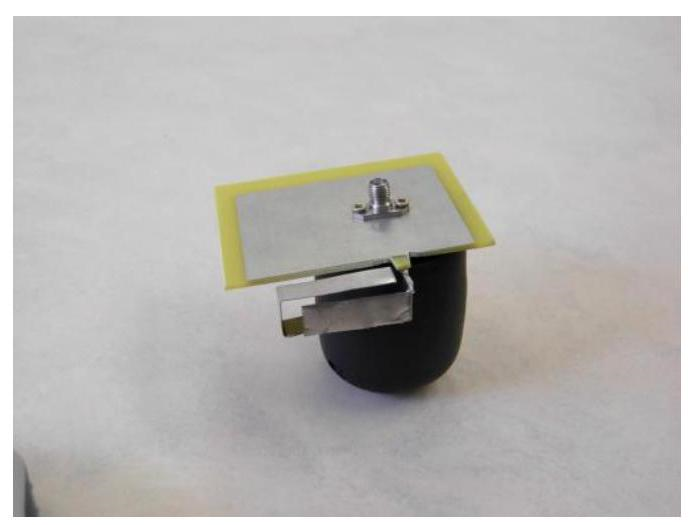}   
        \caption*{(b)}
        \label{fig:(b)}
    \end{subfigure}
   \caption{Wrapped (3D) monopole construction. (a) Front side (b) Back side.}
    \label{fig:6}
\end{figure}

The PCB board used for the antenna construction, shown in \hyperref[fig:6]{Figure 6}, is an MS-FR4 laminate (abbreviation for Flame Resistant 4), which is a thermally stable copper-clad epoxy-glass laminate. The MS-FR4 substrate has a thickness of \(1.6 mm\) and a relative permittivity of \(\epsilon_r=4.7\). The ground plane and patch was constructed from the copper material of the board.\\

\noindent
Two bolt holes of diameter \(2.5 mm\) were drilled into the circuit board at the end of the feed-line in order to secure a pin connector to the board using nuts and bolts. A further hole of \(1.7 mm\) was drilled between the two bolt holes to allow for the pin connector to pass through the substrate and be soldered to the end of the feed-line. The body of the pin connector was earthed to the ground plane.\\

\begin{figure}[h]
\begin{center}
{
\includegraphics[width=0.70\textwidth]{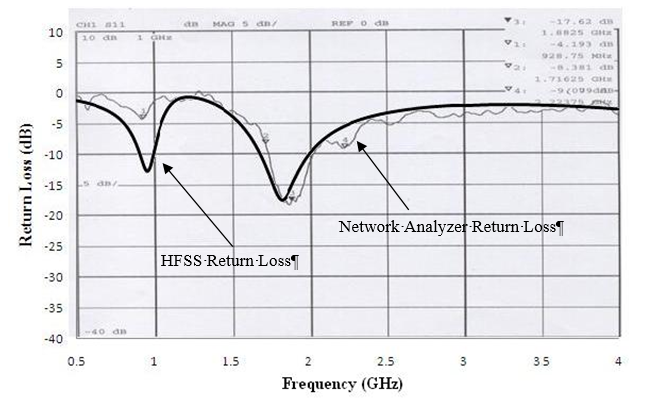}
}  
         \caption{HFSS Return Loss results comparison with experiment.}
         \label{fig:7}
          \end{center}   
\end{figure}

\noindent
\hyperref[fig:7]{Figure 7} shows a comparison of the dual band antenna return loss, generated on a network analyzer, with the theoretical return loss produced by Ansoft HFSS.

\noindent
The overall trend of the HFSS return loss data shows reasonable good agreement with the constructed antenna data. \\

\noindent
In \hyperref[fig:1]{Figure 1 (a)}, it should be noted that DCS band is associated with patch $1$ and the GSM band is associated with patch \(2\). The HFSS DCS frequency band data in \hyperref[fig:7]{Figure 7} gives good agreement with experiment which could be due to the entire length of patch $1$ being perpendicular to the face of the network analyzer during the testing process and therefore less susceptible to stray or reflective radiation from the network analyzer. In the case of the GSM band HFSS results shown in \hyperref[fig:7]{Figure 7}, there is less agreement with experiment than the DCS band HFSS results. This discrepancy could be due to two of the three sides of patch $2$, in wrapped form, being parallel to the face of the network analyzer during the testing stage. Patch $2$ would therefore be more exposed to stray or reflective radiation from the network analyzer leading to a change in return loss gain.

\section{Conclusion}
\label{section5}
In this paper, it has been shown that Ansoft HFSS is a useful tool for designing a dual band monopole antenna to resonate at the frequency bands of interest. A dual band antenna (GSM and DCS) has been fabricated in the laboratory and tested on a network analyser. The experimental return loss data show a good overall trend agreement with HFSS results over the frequency range \(500 \mathrm{MHz}-4 \mathrm{GHz}\). There are some discrepancies in the return loss data which may be attributed to an insufficient length of feedline being used in the construction of the antenna resulting in the monopole being too close to the feed point. Some discrepancy in the return loss data may also occur due to the monopole not being entirely shielded by a ground plane on one side of the substrate. Some reflective radiation can from the network analyser can therefore affect the results.\\

\noindent
More complex shaped wrapped structured monopoles, which contain multiple patch lengths, can be designed in Ansoft HFSS for the antenna to resonate at three or more frequency bands. Fine tuning of the antenna dimensions can be used to adjust the frequency band and to give constant impedance in each band. For improved antenna performance, impedance matching can be achieved for all bands by manipulating the shape of each resonant path without affecting the resonant frequency associated with each path.

\bibstyle{IEEEtran}
\bibliography{anystyle} 
\section*{Competing interests}
The authors declare no competing interests.
\end{document}